\documentclass[12pt]{article}
\usepackage{graphicx}

\begin{document}
\begin{titlepage}
\begin{center}
\centerline{Solid state physics of impact crater formation: further considerations}
\end{center}
\begin{center}
\centerline{V.Celebonovic}			
\end{center}

\begin{center}
\centerline{Institute of Physics,Pregrevica 118,11080 Zemun-Belgrade,Serbia}
\centerline{vladan@ipb.ac.rs}
\end{center}
\begin{abstract}
Impact craters exist on solid surface planets, their satellites and many asteroids.The aim of this paper is to propose a theoretical expression for the product $\rho r^{3} v_{1}^{2}$,where the three symbols denote the mass density,radius and speed of the impactor. The expression is derived using well known results of solid state physics,and it can be used in estimating parameters of impactors which have led to formation of craters on various solid bodies in the Solar System.\footnote{to appear in Rev.Mex.Astron.Astrophys, October 2013}
\end{abstract}
\end{titlepage}





\section{Introduction}
Craters of various sizes have been observed on the terrestrial planets,their satellites and the major part of the asteroids.The study of these craters, and the resulting constraints on the related impactors has become a separate field of research in planetary science,see for example
[8].At the time of this writing, the latest example testifying about the importance of impacts into the Earth and their consequences is the small asteroid which entered the atmosphere over the city of Chelyabinsk in Russia on February 15, 2013. One of the fundamental questions concerning the impactors is what can be concluded about them by combining astronomical data with results of solid state physics.

Recent theoretical work for example [1] has shown that the application of elementary principles of solid state physics to this problem gives physically plausible results in reasonable agreement with those obtained by celestial mechanics. It was assumed there that the material of the target was a crystal lattice,and the calculations were performed {\it per unit volume}. It was assumed in that paper that the condition for the formation of a crater is that the kinetic energy of a unit volume of the impactor has to be equal to the internal energy of a unit volume of the material of the target. The result was an expression for the minimal speed which an impactor of given parameters must have when hitting a target with a predefined set of parameters, in order to form a crater. Obviously,a certain part of surfaces of planets,satellites,asteroids,.. in which impacts occur are granular. For recent theoretical work on "granular impact" see, for example,[3] or [11].   

The aim of the calculation reported here is to take into account the dimensional effects - to consider both the impactor and the crater it forms as objects of finite dimensions. The novelty of the approach discussed in the next section, compared to existing work such as [4],is the generality - it is based on principles of solid state physics,and it can be applied to any solid material. Formation of impact craters is here discussed from the viewpoint of pure solid state physics as the following analogous problem: what kinetic energy of an impactor is needed to produce a hole of given dimensions in a material with a predefined set of parameters? The calculations will be performed under the assumptions that the material of the target is a crystal,that one of the five usual bonding types exists in it and that in the impact the target does not heat up to its melting temperature, so that solid state physics can be applied. For recent work on the problem of heating in impacts see [2]. The expression for the cohesion energy used in the present work is one of the theoretically possible expressions. More details are avaliable, for example, in [7].      

\section{Calculations}

The physical keypoint of the calculation to be discussed in this section is the idea that the kinetic energy of the impactor must be greater than or equal to the internal energy of some volume, $V_{2}$ of the material of the target. The kinetic energy of the impactor of mass $m_{1}$ and speed $v_{1}$ is obviously:
\begin{equation}
	E_{k}=\frac{1}{2} m_{1} v_{1}^{2}
\end{equation}
The internal energy of a volume $V_{2}$ of the target material is equal to the sum of the following three "contributions" : the cohesion energy $E_{C}$,the thermal energy $E_{T}$ and the energy $E_{H}(T)$ required for heating the specimen by an amount $\Delta T$ in the moment of impact: 
\begin{equation}
E_{I}=E_{C}+E_{T}+E_{H}(T)
\end{equation}
The condition for the formation of a crater as a consequence of impact of a projectile into a target is, in general terms, given by
\begin{equation}
	E_{k}=E_{I}
\end{equation}
In order to ensure the stability of matter, the interatomic interaction energy must have the form
\begin{equation}
	E=\frac{-K}{R^{p}}+\frac{C}{R^{n}}
\end{equation}
where $R$ is the inetartomic distance, the first term on the right hand side is attractive and the second one repulsive.It can be shown [5] that the cohesion energy is given by
\begin{equation}
	E_{C}=-\frac{9 B_{0}\Omega_{m}}{pn}
\end{equation} 
where $B_{0} = - V (\partial P/\partial V)_{T} = \rho (\partial P/\partial\rho)_{T}$ is the bulk modulus of a material and $\Omega_{m}$ the volume per pair [5]. The speed of sound waves in a material with pressure $P$ and density $\rho$ is given by 
\begin{equation}
	\bar{u}^{2}= \frac{\partial P}{\partial\rho}
\end{equation}
and therefore
\begin{equation}
	E_{C}=-9\frac{\Omega_{m}B_{0}}{pn}=-9(\frac{\Omega_{m}}{pn})\rho \frac{\partial P}{\partial\rho}=-9\frac{\Omega_{m}\rho\bar{u}^{2}}{pn}
\end{equation}
The analytical form of the function $E_{T}$ depends on the temperature of the solid. It can be shown (for example [7]) that for temperatures 
\begin{equation}
	k_{B}T_{0} >>\frac{\hbar \bar{u}}{a}
\end{equation} 
($\hbar$ is Planck's constant divided by $2\pi$, $\bar{u}$ is the mean speed of sound in the solid and $a$ is the lattice constant),one should use the "high temperature" limit for the thermal energy. Inserting for example $\bar{u} = 5 km/s$ and $a=5\times10^{-6}m$ it follows that $T_{0}>>77 K$. 

The thermal energy $E_{T}$ is given by
\begin{equation}
	E_{T} = 3N \nu k_{B}T D_{3}(x) 
\end{equation} 
where $x=T_{D}/T$, $T_{D}$ is the Debye temperature,$\nu$ is the number of particles in the elementary crystal cell,$N$ is the number of elementary crystal cells in the specimen,and $k_{B}$ is Boltzmann's constant. The symbol $D_{3}(x)$ denotes the $n=3$ case of the Debye function, which is given by 
\begin{equation}
D_{n}(x)=\frac{n}{x^{n}}\int_{0}^{x}\frac{t^{n}dt}{Exp[t]-1}= (n/x^{n}) x^{n}[\frac{1}{n}-\frac{x}{2(n+1)}+\sum_{k=1}^{\infty}\frac{B_{2k}x^{2k}}{(2k+n)(2k)!}]	
\end{equation}
and $B_{2k}$ denote Bernoulli's numbers.
Integrating and retaining the terms up to and including second order leads to: 
\begin{equation}
	E_{T}\cong 3 N \nu k_{B} T [1-\frac{3}{8}\frac{T_{D}}{T}+(\frac{1}{20})(\frac{T_{D}}{T})^2]
\end{equation}
The specific heat $C_{V}$ is given by
\begin{equation}
	C_{V}= \frac{\partial E_{T}}{\partial T} = 3 N\nu k_{B} (D_{3}(T_{D}/T)- (T_{D}/T)D_{3}'(T_{D}/T)) 
\end{equation}
which leads to the following approximate result:
\begin{equation}
	C_{V} \cong 3 N\nu k_{B} [1-(\frac{1}{20})(\frac{T_{D}}{T})^{2}+ (\frac{1}{560})(\frac{T_{D}}{T})^4]
\end{equation}

The energy $E_{H}(T)$ is obviously given by $E_{H}=C_{V}(T_{1}-T)$, where $T_{1}$ is the temperature to which the material of the target heats up in the moment of impact and $T$ is the initial temperature. 

One of the problems consists in knowing the volume $V_{2}$,which is,in first approximation,equal to the volume of the crater. It is known from theoretical work [4] and references given there that the form of a crater is determined by a combination of "gravity scaling" and "strength scaling", where the term "strength" reffers to the material strength of the target. Explicite expressions for the form of craters exist in [4] but they are applicable to 4 material types. In order to simplify somewhat the calculations,and get slightly more general results, it was assumed in the present work that craters have the shape of a half of a rotating elipsoid,with distinct semi axes,denoted by $a$,$b$ and $c$. Physically,$a$ and $b$ denote the semi axes of the "opening" of the crater,while $c$ is the depth. In that approximation,the volume of the crater is  obviously given by
\begin{equation}
V_{2}= \frac{2}{3}\pi a b c
\end{equation}
Inserting the definition of the mass density $(\rho_{1}= \frac{m_{1}}{V_{1}})$ in eq.(1) and assuming further that the impactor has the form of a sphere of radius $r_{1}$,the following expression for the kinetic energy of the impactor is obtained:
\begin{equation}
	E_{k}=\frac{1}{2}m_{1}v_{1}^{2} = \frac{1}{2}\times\frac{4}{3}\pi r_{1}^{3}\rho_{1}v_{1}^{2}=\frac{2}{3}\pi\rho_{1}r_{1}^{3}v_{1}^{2}
\end{equation}
Inserting eqs.(7), (11),(13) and (15) into (3) one gets the final form of the energy condition which has to be satisfied for the formation of an impact crater:
\begin{eqnarray}
3 k_{B} T_{1} N\nu [1-\frac{3}{8}\frac{T_{D}}{T_{1}}-\frac{1}{20}(\frac{T_{D}}{T})^{2}+\frac{1}{10}(\frac{T_{D}^{2}}{T T_{1}})\nonumber\\
+(\frac{1}{560})(\frac{T_{D}}{T})^{4}-\frac{1}{420}\frac{T^{4}}{T^{3}T_{1}}-\frac{3 \bar{u}^{2}\rho\Omega_{m}}{n p \nu k_{B} T_{1}}]
= \frac{2 \pi\rho_{1}}{3}r_{1}^{3}v_{1}^{2}
\end{eqnarray}
The number $N$ is equal to the ratio of the volume of the crater, approximated by eq.(14), and the volume of the elementary crystal cell,$v_{e}$: $N=V/v_{e}$. 
Equation (16) groups known or measurable parameters on the left hand side, while on the right hand side it contains parameters of the impactor. Applying this equation demands the knowledge of the equation of state (EOS) of the material of the target. Generally speaking,regardless of its detailed form, an EOS can be expressed in the following analytical form
\begin{equation}
	P(\rho)=\sum_{i=0}^{\infty}a_{i} (\frac{\rho}{\rho_{0}})^{i}
\end{equation}
where $a_{i}$ are some coefficients, $\rho_{0}$ is the density at some pressure $P_{0}$ and $\rho$ is the density at pressure $P$. It follows that
\begin{equation}
	(\frac{\partial P}{\partial \rho})=\sum_{i=0}^{\infty} (i+1) a_{i+1}(\frac{\rho}{\rho_{0}})^{i}
\end{equation}
A well known example of the EOS of a solid is the Birch-Murnaghan EOS [12]
\begin{eqnarray}
	P(\rho)=\frac{3B_{0}}{2}\left[(\frac{\rho}{\rho_{0}})^{7/3}-(\frac{\rho}{\rho_{0}})^{5/3}\right]\times\nonumber\\
	\left\{1+(3/4)(B_{0}'-4)\left[(\rho/\rho_{0})^{2/3}-1\right]\right\}
\end{eqnarray}
The symbols $\rho_{0}$ and $\rho$ denote the mass density of a specimen under consideration at the initial value of the pressure $P_{0}$ and at some arbitrary value $P$. 

More refined results could be obtained by using analytic approximations to the Helmholtz free energy, from which all the thermodynamic potentials can be derived. This is,for example, the approach used in the ANEOS EOS [13]. However, ANEOS uses a file of up to 40 parameters for the characterization of materials,which complicates its applications.  
\section{A test example}
\label{sec:4}
As a test of the applicability of the procedure discussed in the preceeding paragraph, it was applied to the Barringer crater in Arizona. One of the minerals found in and around the crater is Forsterite - $Mg_{2}SiO_{4}$. The dimensions of the crater are $a=b=1.186 km$ and $c=0.17 km$. Take that the exponents in the expression for the interparticle potential are $p=1$ and $n=9$. Assume furher that $\bar{V}=5 km/s$,which is the mean value of the measured velocity  of seismic waves in the Earth, that the temperature before impact is $T=300 K$ and that the target heats up to $T_{1}=550 K$ at the point of impact. This is safely far from the melting temperature of Forsterite (which is around $T_{M}=1900 C$)  so that solid state physics can be applied. All the parameters of Forsterite exist in the literature, such as [10] or http://webmineral.com. Taking the necessary values,and assuming that the impactor had a radius of $r_{1}=65m$ and density of $\rho_{1}=8500 kg/m^{3}$ one finally gets that $v_{1}\cong41km/s$. 

This value is larger than existing results, such as for example [9] or [6].The discrepancy can be traced to the assumption, made in the calculation reported here, that the target material is $\bf only$ Forsterite. In reality, this is certainly not the case. Taking that $10$ percent of the target is Forsterite,and that the object had a radius of $r_{1}=60m$ would give $v_{1}\cong15 km/s$ , which is much closer to the results of celestial mechanics. 
\section{Discussion}
In this paper we have presented a simple procedure which gives the possibility of estimating the value of the product of the density and radius of the impactor and the speed of impact: $\rho_{1} r_{1}^{3} v^{2}$ in terms of various material parameters of the target and the impactor. The expression for this product has been derived using basic principles of solid state physics, without any special assumption(s) about the materials. In (at least some of the) terrestrial applications the density of the impactor and the inclination of the trajectory can (in principle) be measured or estimated, which gives the possibility of estimating the impact speed if the value of $r_{1}$ can somehow be estimated. 

The shape of the craters has been approximated as a half of a rotating elipsoid,regardless of the composition of the material of the target. As a first approximation, the speed of the seismic waves has been "put in" instead of being calculated from the equation of state of the material of the target. Calculating this speed would have demanded the precise knowledge of the chemical composition of the target materials,and of $B_{0}$ and $B_{0}'$ for them.  
The impact of a projectile into a target leads to heating,and possibly melting and even vaporisation of the target around the impact point. The thermodynamic result of an impact depends on the heat capacity of the material of the target and on the kinetic energy of the  impactor. The heat capacity can be measured, or theoretically estimated, assuming prior knowledge of the chemical composition of the target. Knowledge of the heat capacity is vital for estimates of the temperature to which the target heats in the impact.Therefore, the question is which impacts lead to melting and vaporisation of the material of the target,and which just provoke heating of the target. Even if the material of the target gets partially melted and as such flows away, the dimensions of the crater formed by the impact are determined by the volume of the material pushed away in the impact. This volume is,in turn, determined by the ratio of the kinetic energy of the impactor to the internal energy of some volume of the target. Some details of this problem have been discussed in [2].

The general conclusion of work reported here is that by using simple well known results of solid state physics and a given appoximation of the shape of impact craters, it becomes possible to estimate the value of the product $\rho_{1} r_{1}^{3} (v \cos \theta)^{2}$. This in turn can be used to draw conclusions about the impactors which made various craters on the solid surfaces of objects in the planetary system.  

\section{Acknowledgment} 
This paper was prepared within the research project 174031 financed by the Ministry of Education Science and Technological Development of Serbia. I am grateful to the referee for helpful comments and to Dr. Jean Souchay from Observatoire de Paris for correspondence .

\end{document}